# A Geologically Robust Procedure For Observing Rocky Exoplanets to Ensure that Detection of Atmospheric Oxygen is an Earth-Like Biosignature

**Carey M. Lisse[1,*], Steven J. Desch[2], Cayman T. Unterborn[2], Stephen R. Kane[3], Patrick R. Young[2], Hilairy E. Hartnett[2,4], Natalie R. Hinkel[5], Sang-Heon Shim[2], Eric E. Mamajek[6], Noam R. Izenberg[1]**

[1] Applied Physics Laboratory, Johns Hopkins University, 11100 Johns Hopkins Rd., Laurel, MD 20723. carey.lisse@jhuapl.edu, noam.izenberg@jhuapl.edu

[2] School of Earth and Space Exploration, Arizona State University, PO Box 871404, Tempe AZ 85287-1404. steve.desch@asu.edu, cunterbo@asu.edu, patrick.young.1@asu.edu, h.hartnett@asu.edu

[3] Department of Earth and Planetary Sciences, University of California, Riverside, 900 University Ave., Riverside CA 92521. skane@ucr.edu

[4] School of Molecular Sciences, Arizona State University, PO Box 871604, Tempe, AZ 85287-1604.

[5] Southwest Research Institute, San Antonio, TX, USA 28510. natalie.hinkel@gmail.com

[6] Jet Propulsion Laboratory, 4800 Oak Grove Drive MS 321-100, Pasadena, CA 91109-8099. mamajek@jpl.nasa.gov, emamajek@gmail.com

27 Pages, 1 Figure, 0 Tables

Key Words: **Exoplanet systems; Exoplanet structure; Stellar astronomy; Astrobiology; Biomarkers;**

**\*Corresponding author: carey.lisse@jhuapl.edu, (240) 228-0535**

Full Title:  **"A Geologically Robust Procedure For Observing Rocky Exoplanets to Ensure that Atmospheric Oxygen is an Earth-Like Biosignature"**

Proposed Running Title:  **"Rocky Exoplanet Oxygen Biosignatures"**

Please address all future correspondence, reviews, proofs, etc. to:

Dr. Carey M. Lisse

Planetary Exploration Group, Space Exploration Sector

Johns Hopkins University, Applied Physics Laboratory

11100 Johns Hopkins Rd

Laurel, MD 20723

240-228-0535 (office) / 240-228-8939 (fax)

Carey.Lisse@jhuapl.edu

# Abstract

In the next decades, the astrobiological community will debate whether the first observations of oxygen in an exoplanet's atmosphere signifies life, so it is critical to establish procedures now for collection and interpretation of such data. We present a step-by-step observational strategy for using oxygen as a robust biosignature, to prioritize exoplanet targets and design future observations. It is premised on avoiding planets lacking subaerial weathering of continents, which would imply geochemical cycles drastically different from Earth's, precluding use of oxygen as a biosignature. The strategy starts with the most readily obtained data: semi-major axis and stellar luminosity to ensure residence in the habitable zone; stellar XUV flux, to ensure an exoplanet can retain a secondary (outgassed) atmosphere. Next, high-precision mass and radius information should be combined with high-precision stellar abundance data, to constrain the exoplanet's water content; those incompatible with $< 0.1 \mathrm{wt\%}$ $H_2O$ can be deprioritized. Then, reflectance photometry or low-resolution transmission spectroscopy should confirm an optically thin atmosphere. Subsequent long-duration, high-resolution transmission spectroscopy should search for oxygen and ensure that water vapor and $CO_2$ are present only at low ($\sim 10^2$-$10^4$ ppm levels). Assuming oxygen is found, attribution to life requires the most difficult step, acquisition of a detailed, multispectral light curve of the exoplanet, to ensure both surface land and water. Exoplanets failing some of these steps might be habitable, even have observable biogenic oxygen, but should be deprioritized because oxygen could not be attributed unambiguously to life, and life therefore would not be *detectable* on such planets. We show how this is the case for the solar system, the 55 Cnc system, and the TRAPPIST-1 system, in which only the Earth and TRAPPIST-1e successfully pass through our procedure.

# I.     Introduction

The search for life on planets around other stars is one of the grand scientific challenges of the 21$^{st}$ century. Although rocky exoplanets around Sun-like stars have been discovered only in the last two decades, technologies and strategies are being developed now to look for life on these planets, with the search likely to reach fruition in the next few decades. At the 2017 Habitable Worlds Conference in Laramie, Wyoming, a poll of the audience of exoplanet astronomers and astrobiologists yielded a median estimate of 2050 for when definitive signs of life on an exoplanet would be discovered (Hannah Jang-Condell, personal communication). One approach being adopted by the astronomical community is to find putative biosignature gases—especially oxygen and methane— in the atmospheres of exoplanets, through infrared transmission or emission/reflectance spectroscopy (Domagal-Goldman *et al.* 2016, Arney 2019).

We focus here on the plan to measure oxygen on a planet **harboring life as we know it on Earth**. If Earth today were treated as an exoplanet, oxygen—either in the form of $O_2$ or its derivative, $O_3$—would be among the most readily seen and best understood of the biosignature gases we could observe in its atmosphere (Meadows *et al.* 2018). We recognize that an exoplanet could be habitable, even life-bearing, and yet not export gases to an atmosphere; a Solar System analogy could be Europa, if it harbors life under its ice shell. Even on the Earth, biogeochemistry was not dominated by oxygenic photosynthesis for billions of years - Proterozoic Earth has been suggested to have had extremely low atmospheric $O_2$ levels as recently as 0.8 Gya (Planavsky *et al*. 2014). We also recognize that many additional biosignature gases (e.g., $CH_4$, $N_2O$, etc. per Seager *et al.* 2016) have been identified and that these species should be considered in future papers using the same framework we provide here.

It is also important to acknowledge that geochemical cycles on an exoplanet are likely to differ from Earth's, which entails important ramifications for use of oxygen as a biosignature. One of the most powerful determinants of oxygen as a useful biosignature is the water content of an exoplanet. Planets with water contents much greater than Earth appear more prone to false positives for oxygen (e.g., Luger & Barnes 2015), but desiccated planets may also be susceptible to $O_2$ false positives (Meadows *et al.* 2018). Thus, a comprehensive examination of a planet's geochemical cycles, including the particular effects of water content, is needed to interpret whether oxygen is a true biosignature on various exoplanets and to help plan future observations.

We draw on results in a companion paper (Glaser *et al.* 2020*)* that atmospheric oxygen is a useful biosignature on rocky, Earth-like (i.e., radius < 1.5 $R_E$) exoplanets only if they have both surface water and land, or exposed continental rock. As explained below, this is because exoplanets without continents are expected to have biological oxygen production rates indistinguishable from abiotic processes, so it will be challenging to determine whether such planets actually host life. Understanding these production rates is critical, because while astronomers can constrain the abundance of biosignature gases like $O_2$ through spectral retrieval techniques, inferring life depends on determining the rate of production of any given biosignature, which in turn requires understanding additional context about the planet.

The Glaser *et al.* 2020 finding is premised on the way oxygen produced by life is exported to an atmosphere. In using oxygen as a biosignature, it is implicitly assumed that the atmosphere is in steady state and destruction balances production. We also take as a given that the $O_2$ destruction chemistry is similar to Earth's, with $O_2$ consumed by reactions with reduced outgassed species (e.g., $H_2S$, $H_2$, CO, $CH_4$) and reduced minerals (e.g., $FeS_2$). The lifetime of $O_2$ in modern Earth's atmosphere is ~1.7 Myr, and the 3.3 x $10^7$ Tmol of $O_2$ in modern Earth's atmosphere is drawn down by these reactions at a rate of 20 Tmol/yr (Catling 2014). Only if a lifetime against reduction could be derived for an exoplanet could an inferred mass of oxygen in an atmosphere be converted into a destruction rate and therefore a production rate. Determination of this lifetime is one way in which geochemistry matters to use of oxygen as a biosignature.

Assuming a production rate is obtained, the next question is whether it can be assigned a purely geological, non-biological origin or can be attributed to life. This requires a theoretical comparison of production rates on hypothetical exoplanets. On modern Earth, life produces $O_2$ at a rate $\sim 10^4$ Tmol/yr by oxygenic photosynthesis that consumes water and $CO_2$ to produce carbohydrates and $O_2$. It simultaneously consumes $O_2$ at $\sim 10^4$ Tmol/yr via respiration that combines carbohydrates and $O_2$ to produce $CO_2$ and water. Due to a slight imbalance in these rates associated with burial of organic carbon (Catling 2014), this leads to 20 Tmol/yr net export to the atmosphere. This process of carbon burial inevitably sequesters other bioessential elements, most importantly phosphorus. The export of $O_2$ to the atmosphere is directly proportional to the flux of bioavailable phosphate from weathering of apatites on continents.

On planets with submerged continents, i.e., no subaerial weathering, the flux of phosphate is reduced due to changed pH and erosion rates, by three orders of magnitude, to 0.02 Tmol/yr (Glaser *et al.* 2020). This is comparable to abiotic production rates of $O_2$ by photolysis of $H_2O$, followed by hydrogen escape. An exoplanet with as little as 0.1wt% water on its surface—what we term a 'pelagic planet'—would have no exposed land and no subaerial weathering of continents (Cowan & Abbot 2014). On such a planet, it would be difficult to know the rate of reduction of oxygen and turn an observed abundance into a production rate; but assuming this could be done correctly, the resulting inferred production flux would be $\sim 0.02$ Tmol/yr, with considerable uncertainty. Even if this production of $O_2$ were actually due to oxygenic photosynthetic life, it would be impossible to rule out with confidence an origin in abiotic processes. Therefore, regardless of the actual mass of $O_2$ in the atmosphere or how hard or easy it is to observe, and separate from issues of outgassing of reduced species, the geochemical cycles on a planet with > 0.1 wt% water preclude using $O_2$ as a biosignature. We could directly observe the byproducts of life, and yet life would not be *detectable*.

Before the *Transiting Exoplanet Survey Satellite* (*TESS*, Ricker *et al.* 2015), there were very few, if any, known nearby transiting rocky exoplanets with atmospheres that could be measured using state-of-the-art 2020 transmission or emission spectroscopy. To detect these biogases using transit spectroscopy would require at least the sensitivity of the upcoming *James Webb Space Telescope* (*JWST*) mission, if not the Flagship-class observatories (*LUVOIR*, the *Large UV/Optical/Infrared Surveyor*; *HabEX*, the *Habitable Exoplanet Observatory*; and *Origins*, the *Origins Space Telescope*) currently under study.

Until such missions launch in the 2030's, exoplanet characterization would be carried out by *JWST*. During its nominal lifetime (2021-2026), it appears possible to use *JWST* to characterize the atmospheres of ~5 known nearby transiting rocky exoplanets (Morley *et al.* 2017; Lustig-Yaeger *et al.* 2019). The low number of known measurable HZ rocky transiting exoplanets is an observational effect, not a physical one; there are numerous faint and distant low-mass, rocky Kepler/K2 exoplanets in their stars' habitable zones (HZs) per the NASA Exoplanet Archive[1]. *TESS* has already begun to find many more nearby measurable transiting rocky exoplanets (Cloutier *et al.* 2019, Gilbert *et al.* 2020). It has been estimated that *TESS* could find dozens in their stars' HZs (Sullivan *et al.* 2015), although 1-3 is more likely (Barclay *et al.* 2018). By ~2030 the *CHaracterizing ExOPlanet Satellite* (*CHEOPS;* Deline *et al.* 2020) and *PLAnetary Transits and Oscillations of stars* (*PLATO;* Rauer *et al.* 2014) spacecraft missions, and the ground based SPECULOOS red dwarf survey (Delrez *et al.* 2018) will each observe hundreds to thousands of planets, a noteworthy fraction of which will be rocky planets, and a percentage of these will be in the habitable zone (i.e. tens of rocky HZ planets *in toto*). Other ground-based ELTs characterizing Earth-like exoplanets may also come on-line if current plans come to fruition (e.g., Quanz *et al.* 2015 or Snellen *et al.* 2015 using the under construction 39m E-ELT with the proposed METIS instrument). A strategy will be needed to prioritize these new planets for follow-up observations. A clear, rigorous filtering triage scheme must be adopted in order to focus first on the exoplanets most likely to give an unambiguous signature of life, if it exists.

The difficulties associated with finding suitable planets and performing transmission spectroscopy are considerable. During the nominal mission lifetime of *JWST*, it may only be able to acquire spectra for exoplanets likely to yield the highest signal-to-noise (e.g., the TRAPPIST-1 planets). Target selection may not be relevant before the 2030s. Nevertheless, our findings are important in the near term, especially with respect to the interpretation of those spectra. It is necessary to know now that confident assignment of $O_2$ to life requires confirmation of both surface land and water on an exoplanet, so that missions like these can be designed and yield needed data by the time they launch.

---

[1] https://exoplanetarchive.ipac.caltech.edu/

Requiring life to be *detectable* using $O_2$ biosignatures demands a different observational strategy to help prioritize exoplanet targets for expensive observations, and to help design future astronomical missions. To aid in interpretation of acquired spectra, selection of exoplanets for expensive observations to acquire transmission spectra, and to help in the design of future missions, we present here a framework for carrying out observations. In this *Letter* we outline a step-by-step process (**Figure 1**), premised on using more easily obtained observational data to rank exoplanets around FGKM stars for more difficult follow-up observations. Our approach resembles triage schemes promulgated in the HabEx, LUVOIR, and Origins Mission Study Final Reports (Fischer *et al.* 2019, Meixner *et al.* 2019, Gaudi *et al.* 2020), but this work, developed in parallel, adds critical multi-wavelength, multi-telescopic measurements and detailed geochemical/geophysical modeling steps to the schema. Ultimately, the goal is to maximize the likelihood that a planet could eventually be demonstrated to not only have $O_2$ in its atmosphere, but could also be demonstrated to have land and water on its surface, so that the $O_2$ would be a reliable and defendable biosignature.

## II.    Too Much Water Obscures the Signs of Life

Life as we know it requires water, and water is equated with habitability (Kasting & Catling 2003, Lammer *et al.* 2003). It is easy, but fallacious, to assume that Earth is water-rich and we should therefore look for life on water-rich planets; i.e., the more water, the better. From a bulk planet perspective, Earth is in fact quite water-depleted, with surface oceans that make up only 0.02% of Earth's mass (0.02 wt%). And, as Glaser *et al.* (2020) conclude, searching for life becomes increasingly more difficult as the water fraction exceeds 0.1wt%. Just 5 oceans' worth of water on the surface of a 1 $M_E$, 1 $R_E$ planet is sufficient to submerge all continents, assuming standard topography (Cowan & Abbot 2014). Lack of subaerial weathering can reduce the flux of bioavailable phosphate by about 3 orders of magnitude. On an Earth-like planet with 50 oceans (just 1wt% bulk $H_2O$) any continents and geochemical cycles would take place under a thick (~100 km) high-pressure ice mantle that would cut off chemical communication between the rocky planet and the oceans (Leger *et al.* 2004; Fu *et al.* 2010; Noack *et al.* 2016). On an Earth-like planet with just 2 wt% bulk $H_2O$, silicate melting and outgassing would be suppressed by the high pressures, effectively shutting off geochemistry altogether (Kite *et al.* 2009). Only if a planet had < 0.1 wt% $H_2O$ could we be sure that the biogeochemical cycles were sufficiently like Earth's to use $O_2$ as a biosignature gas.

This being said, it is important to state that Earth is the only known planet with liquid water on its surface (*i.e.*, a pelagic planet) and the only known planet with life as we know it teeming on its surface. It is also important to recognize that a pelagic planet with, say, > 0.1 wt% water, equivalent to > 5 oceans' worth of water on an Earth-mass planet, could still be habitable and have the same geochemical cycles as Earth (Glaser *et al.* 2020). We could even observe the $O_2$ generated by any oxygenic photosynthesizing organisms in the planet's deep oceans. The biosphere could generate (and consume) $O_2$ at great rates, perhaps even Earth-like rates ~$10^4$ Tmol/yr. But the net export of $O_2$ to the atmosphere would be limited to ~0.02 Tmol/yr. Despite being quite habitable, such a planet would not be suitable for looking for life, at least using $O_2$. For these reasons, not only must an abundance of $O_2$ be found in an exoplanet's atmosphere, surface land and water also must be confirmed, to infer production rates and use $O_2$ to detect life.

## III. An Observational Procedure for Observing Exoplanets

Current techniques using planetary mass and radius can identify planets with > several wt% $H_2O$, but these should be deprioritized for observations because oxygen would not be a reliable biosignature on them. **Although life as we know it requires water, we should actually search for life on planets with less water than is currently detectable.** Filtering observations should be undertaken in the order laid out in our procedural floswchart, which are ordered by degree of resources needed to make the measurement + assessment (Figure 1). Here we expand on these steps.[2]

**<u>Step 1</u>:** Determine as precisely as possible exoplanet masses (M) and radii (R), and host star stellar parameters: surface gravity (log g); effective temperature ($T_{eff}$); age; mass ($M_\star$); luminosity ($L_\star$); and the proxy for overall metals abundances, [Fe/H]. Planetary radii will be most precisely determined from precision transit photometry using well-characterized host stars. Masses should be derived from radial velocity (RV) or astrometry measurements, if possible, but potentially are much more precisely determined from transit timing variations (TTVs) for exoplanets in multi-planet, transiting systems (e.g., TRAPPIST-1; Gillon *et al.* 2017; Grimm *et al.* 2018).

Use these quantities to triage exoplanets, prioritizing those that are rocky exoplanets with Earth-like atmospheres, currently in the HZs of their main-sequence dwarf stars. Planets with radii > 1.5 $R_E$ can be deprioritized, because they are very likely to have thick $H_2$/He atmospheres (Weiss & Marcy 2014, Rogers 2014, Fulton *et al.* 2017), complicating the detection of $O_2$ and implying in any case radically different geochemistry. Planets with radii below some lower limit of ~0.6 $R_E$ may not be able to retain atmospheres, although the lower bound is sensitive to many factors (Zahnle and Catling 2017).

More restrictively, use mass-radius relations (e.g., Dorn *et al.* 2015, Zeng *et al.* 2016) and deprioritize those planets with radii larger than is expected for a rocky exoplanet of the measured mass. For a 1 $M_E$ planet, this would be ~1.1 $R_E$.

Deprioritize exoplanets outside their star's HZ (e.g., Kopparapu *et al.* 2014), as well as those planets that have not spent a minimum time (e.g., ~1 Gyr) in the HZ (Truitt

---

[2] N.B. - Much of steps 1–4, and parts of some of the subsequent steps, are not specific to planets with oxygen-producing biosignatures and are useful for choosing habitable planets that can host life in a more general sense. Also, several of the later steps in our schema are currently very difficult to do with existing telescopes, and we have done our best to extrapolate to future observing capabilities. However, some re-ordering of the flowchart may be necessary in $10 – 20$ years' time – *e.g.*, if in the future it becomes easier to do detailed transit spectroscopy than planetary reflectance photometry, or to study other constraining biosignature gases, then the spectroscopy steps should be moved before the reflectance steps. The general gateway logical structure of our schema would not change, only the order of Steps 5 to 8.

*et al.* 2015), so that the planet has had a sufficiently long time to develop oxygenic photosynthesis, or similar time-dependent habitability considerations. This requires determining a host star's minimum possible age to 0.1 Gyr accuracy, coupled with stellar and HZ modelling that relies on the measured stellar and planetary parameters.

**Step 2:** Determine the current X-ray/ultraviolet (XUV) fluxes and infer the past XUV fluxes of host stars. All-sky survey flux values are available from the 1990's era *ROentgen SATellite* (*ROSAT*) All-Sky Survey (RASS) catalog for $\sim 10^5$ of the brighter stellar sources of all ages and stellar types, with a newer deeper *eROSITA* all-sky x-ray photometric survey just begun (Predehl 2014, 2017). New, detailed measurements of individual systems, at least down to energies 0.1 keV, can be obtained using data from the *Chandra* (Weisskopf *et al.* 2000) or *X-ray Multi-Mirror (XMM)-Newton* (Struder et al. 2001) missions (e.g. Lisse *et al.* 2017). Past activity can be inferred from the host's present day XUV behavior extrapolated backwards in time using temporal trending of stellar x-ray behavior (e.g. Lammer et al. 2003, Ribas et al. 2005, Telleschi et al. 2005, Güdel 2007, Guinan & Engle 2007, Osten & Wolk 2015) constrained by the observed statistical XUV behavior (e.g Suchkov & Schultz 2001, Suchkov et al. 2003, Schmitt & Liefke 2004, Güdel 2009, Testa 2010) of the host star's stellar type , which requires the precision stellar parameters found in Step 1.

Presuming that exoplanets typically form with thick $H_2$/He atmospheres accreted from their protoplanetary disks (Stokl *et al.* 2015), it is necessary that past stellar activity has been sufficient to strip the planets of these primary atmospheres. Depending on whether an Earth-mass planet formed from smaller planetary embryos (e.g., Earth probably formed from the merger of two embryos, 0.4 $M_E$ and 0.6 $M_E$: Canup 2012; Desch & Robinson 2019) or directly from the protoplanetary disk as in models of pebble accretion (Chambers 2016 and references therein), planets should be born with $H_2$/He atmospheres with pressures $\sim 1$-$10^4$ bar, or up to several $\times\ 10^{-3}$ $M_E$ of $H_2$/$H_e$. The criterion is that the integrated XUV heating over time exceed roughly 40% of the planet's gravitational binding energy (Lopez & Rice 2018), or about $10^{39}$ erg for a 1-$M_E$ planet. This is not very restrictive, and is expected that planets with radii < 1.5 $R_E$ would have lost their primary atmospheres.

It is also critical that the total integrated XUV flux on a planet from its birth through the present day not exceed the threshold necessary for retention of its secondary atmosphere. For a 1-$M_E$ planet in the habitable zone of a G star, this is less than an order of magnitude above that experienced by Earth, ~several $\times 10^{46}$ erg (Zahnle and Catling 2017).

**Step 3:** Obtain precise host-star elemental abundance ratios. Mass-radius relationships show that if planets are similar to their host stars in composition, the radius of a rocky exoplanet even of a fixed mass can vary significantly. To better exclude those planets with even ~1 wt% $H_2O$ on their surfaces, we want to constrain ratios bearing on gross mantle mineralogy (e.g., Si/Mg), melting relations (Na/Mg and Al/Mg), and radiogenic heat fluxes (U/Mg, Th/Mg, K/Mg). For example, variations across the range of observed stellar compositions (Fe/Mg = 0.4 to 1.5) lead to 20% variations in exoplanet bulk mass and density (Unterborn & Panero, 2019). Equivalently, for a given mass, the radius could vary by 6%. In principle, to exclude a 1 wt% surface abundance of $H_2O$, the stellar Fe/Mg ratio must be constrained to 5% accuracy, i.e., 0.02 dex (Hinkel & Unterborn 2018).

Other host star rocky element abundance ratios are important to obtain, to understand how geochemical cycles could present on an exoplanet. The Si/Mg ratio determines water storage in a planetary mantle and partitioning of water between the surface and mantle. As a result, significant variations in mineralogy accompany shifts in this ratio if >10% different from Earth. Exoplanet prioritization would benefit from constraining Si/Mg in the host star to better than 10%, i.e., 0.04 dex. (NB: the Earth differs from the Sun in this ratio, but only by 20%, i.e., 0.08 dex). The rheology of the mantle, to which convection is sensitive, could also depend sensitively on this ratio. Melting curves, and therefore the thickness of a lithosphere or the pressures at which degassing occur, depend on Si/Mg but also Na/Mg and Al/Mg ratios, which should be similarly constrained.

The exchange between the mantle and surface depends greatly on whether a planet has plate tectonics or a stagnant lid. In turn, the style of surface depends on the vigor of convection and therefore heat flux, which is tied to internal radiogenic heating levels (note that we are assuming here that for > 1 Gyr old planets, any heating due to giant impacts during final planetary aggregation, later surface patina emplacement, or tidal orbital locking has long since equilibrated). Exactly how these elemental abundances affect planetary geodynamics and subsequent atmospheric degassing is still a very active area of research across the geosciences (e.g., Foley & Syme 2018); but these studies highlight the importance of measuring the U/Mg, Th/Mg, and especially the K/Mg (technically, only radiogenic $^{40}K$ = 0.012% of K on Earth is important, but we can use K as a proxy for $^{40}K$) ratios. Even presuming a stagnant lid regime, the lifetime of degassing is sensitive to these ratios, which should be measured to about 10%, or 0.04 dex (Unterborn *et al.*, submitted). In most respects, the closer these ratios are to Earth-like, the more likely the planet is to degas volatiles into their atmospheres and to emplace minerals with bioessential elements (e.g., P) at Earth-like rates. This in turn will increase the predictive power of models of the exoplanet's geochemistry, so the highest priority should be given to exoplanets whose host-star abundance ratios generally match the Sun's to within < 0.1 dex (*e.g.* Bedell, Bean *et al.* 2018), which is approximately 35% of all nearby sun-like stars (Hinkel *et al.* 2017).

To a lesser extent we want to measure the harder-to-obtain stellar abundances of the bioessential volatile elements N, P, S, and K. These are required elements for all oxygenic photosynthetic life on Earth. To apply information about these elements to models of biospheres, we must first determine the abundances of these elements relative to major-rock forming elements (e.g., N/Si, P/Si, K/Si). These molar ratios are themselves derived from the distributions among stars and their correlations with each other (Hartnett, Hinkel, & Young 2019). Because of their volatility, N and P especially can be fractionated with respect to Si during the planet-formation process, so the molar ratios at a planet's surface may not match those in the star; but correlations between N/Si and P/Si would be useful. Unfortunately, there is a dearth of stellar abundances for these elements in the *APOGEE* survey (Smith *et al.* 2013), the *GALAH* survey (de Silva *et al.* 2015) and the *Hypatia Catalog* (Hinkel *et al.*

2014). This is because they are difficult to measure in stars: most high-resolution spectrographs focus on lines in the optical, but many of the strongest lines for these elements exist in the ultraviolet or infrared and are obscured or contaminated by Earth's atmosphere. For example, the simultaneous measurement of N, P, and Si abundances has been performed in only 51 stars (0.8% of the *Hypatia Catalog*). Due to the small numbers, it is not possible to make inferences about the correlations of N/Si vs. P/Si.

**Step 4:** Use the precise host-star abundance measurements to refine the planetary interior modeling, to better constrain the surface water content. Such modeling must include detailed mineralogy as a function of depth, and must allow for distinctly non-Earth-like compositions. Modeling should also include new equations of state for rock-water composites at high pressures, to assess the probability that the observed mass and radius of the exoplanet is consistent with no measurable (i.e., < 0.1 wt%) water.

Simple mass-radius scaling models (e.g., Zeng & Sasselov 2013; Dorn *et al.* 2015; Zeng *et al.* 2016; Zeng & Jacobsen 2017) do not accurately predict the interior structure for planets, including the Earth (Unterborn *et al.* 2016, Unterborn & Panero 2019) and more advanced calculators must be employed that self-consistently calculate the mineralogy of an exoplanet from the elemental abundances provided in Step 3 (e.g. Dorn *et al.* 2015, Unterborn *et al.* 2018). Realization will require additional modeling and experimental work: while many experiments constraining the equation of state for Fe-cores (e.g., Smith *et al.* 2018; Wicks *et al.* 2018) are nearing the pressures expected within planets with radii up to 1.5 $R_E$ (Unterborn and Panero 2019), few experiments have been performed for mantle silicates at these relevant pressures, and only some computational work for silicates above the pressure and temperature conditions expected inside Earth. These computations suggest dissociation of mantle silicates into constituent oxides (FeO, MgO, $SiO_2$) at high pressures (e.g., Umemoto & Wentzcovitch 2011; Umemoto *et al.* 2017), although the conditions required for this reaction are more relevant to mini-Neptunes than super-Earths (Unterborn and Panero 2019). In general, a family of mass-radius curves should be generated, using planet compositions matching the stellar compositions (including Fe/Mg, etc.), which should be determined as precisely as possible.

The exoplanets that should be prioritized are those with the smallest probability of their combination of (imprecisely) measured mass and radius placing them above the family of mass-radius curves that indicate rocky composition. Those that do lie above these curves can be inferred to have abundant volatile layers of atmospheric $H_2$/He or $H_2O$. By assessing the ability of XUV flux to remove an $H_2$/He atmosphere (Step 2), it is more likely that water comprises the volatile envelopes on exoplanets that are the focus of Step 4. Although it may not be possible to rule out surface water fractions as low as 0.1 wt%, even constraining the surface water content to a fraction of 1 wt% would be significant.

**Step 5:** If the host star's flux allows it (M-stars with very low UVIS emission are unlikely to create enough planetary UVIS signal to be detectable), obtain "fast" 3-color UVIS transmission/reflectance photometry of the exoplanet. Use the results of Trauger & Traub (2007) and Crow *et al.* (2011) to determine if the terrestrial exoplanet is Earth-like in hosting a bright bluish atmosphere due to water vapor + oxygen + nitrogen. This is opposed to a pale-blue, $H_2$/He/$CH_4$-dominated atmosphere (like Uranus or Neptune), a dense, yellowish-white atmosphere dominated by $CO_2$ and clouds (like Venus), or no atmosphere at all. Only if the planet's B/V/R (e.g., 350/550/850 nm bandcentered) colors are similar (to within ± 40%) of Earth's should the object be prioritized for the much more expensive (in terms of observing time) next step (Krissansen-Totton *et al.* 2016, Izenberg *et al.* 2018, Stickle *et al.* 2019). Note that this step will also remove Earth-like planets with clouds so thick that they push the atmosphere's nominal 350/550 and 850/550 nm ratios towards unity (Krissanen-Totton *et al.* 2016), but these planets will have unobservable surfaces (see next step) and thus indeterminant biosignatures.

**Step 6**: Perform low-spectral resolution (R = 10 - 100) exoplanet transmission spectroscopy (for transiting planets) or direct imaging reflectance/emission spectrophotometry (for wide-orbit planets) to determine the planet's color. A lack of significant variation with wavelength (as for GJ1214b: Kreidberg *et al.* 2014) would indicate either the lack of an atmosphere or the presence of optically thick hazes. Since observations of the lower atmosphere and, eventually, the planetary surface are demanded, only those exoplanets with significant (~10%/100 nm from 400 to 1000 nm) flux variation with wavelength should be further characterized.

**<u>Step 7</u>:** Perform high signal-to-noise, moderate resolution (e.g. S/N 10 − 20, R = 140; Feng *et al.* 2018) and/or ultra-high resolution (e.g. R ~ 100,000; Snellen *et al.* 2015) spectroscopic observations in transmission (for transiting planets) or reflected light (for wide orbit planets) to search for oxygen and other important spectral features. This step is likely to take $10^2 - 10^3$ hours of giant-class (10-30m) telescope time (Krissansen-Totten *et al.* 2016; Kopparapu *et al.* 2018; Izenberg *et al.* 2018, Stickle *et al.* 2019), but biosignature $O_2$ at concentrations of tens of % (i.e., ~ 0.1 bar partial pressure), and $CH_4$ at ppm levels, could be detected (Reinhard *et al.* 2017, Krissansen-Totten *et al.* 2018, Olson *et al.* 2018). Using $O_3$ as a proxy for $O_2$, its main atmospheric source, can extend $O_2$ detectability down to the % range by utilizing ozone's very strong UV and mid-IR absorption features. Simultaneous detection of reduced $CH_4$ and oxidizing $O_2$, which are in chemical disequilibrium with each other, would be a highly robust indicator of complex, established life producing steady state metabolic products (e.g., Schwieterman *et al.* 2018). We note that these detections are not possible for young, archaean biospheres, which are unlikely to have evolved detectable oxygenic photosynthesis and converted the bulk of their primordial terrestrial planet $CO_2$ into carbonate + $O_2$ (e.g., Krissansen-Totten *et al.* 2018). Approximately 0.1% $H_2O$ atmospheric vapor should be found to signal the presence of liquid water on the surface (Betermieux & Kaltenegger 2014), but not more (Glaser et al. 2020), and the water vapor must be restricted to the troposphere, as planets in runaway moist greenhouse stages will have significant amounts of stratospheric water (Kopparapu *et al.* 2013, Luger and Barnes 2015). However, if $CO_2$ is present at ~1 bar levels, it indicates a breakdown in the planet's carbonate-silicate-nitrate-phosphate geological cycles; thus supplies of bio-essential elements N and P will also likely be greatly reduced (Glaser *et al.* 2020). Such planets should be de-prioritized. If CO is spectroscopically found in large abundances (> 0.01 bar), then any detected $O_2$ is likely dominated by abiotic $CO_2$ photolysis (Schwieterman *et al.* 2016). Similarly, if $O_2$ is found at 10 − 100 bar levels, it would indicate a massive abiotic signal from ocean hydrolysis (Schwieterman *et al.* 2016, Meadows *et al.* 2018s), making any biotic $O_2$ signature impossible to measure, thus also disfavoring the planet for further observations.

**Step 8:** Obtain the planet's optical reflectance light curves to search for evidence of continents and oceans. **If and only if the previous steps have indicated the presence of an *imageable* planet with high detectability of life (if it exists) should attempts be made to measure the optical reflectance light curve.** This step is likely to take $10^2 - 10^4$ hrs of giant class (10-30m) telescope time, and many exoplanet orbits (Lustig-Yaeger *et al.* 2018). Photometric observations at continuum wavelengths of reflected light will be used to identify continent-sized regions of land and open water. Principal component analysis of the time-varying reflected light in various filters has allowed detection of land and oceans on Earth from space, and could be used to identify patches of land, ocean, and vegetation on an exoplanet (Cowan *et al.* 2009; Fujii *et al.* 2010). Open water could be independently verified if there was evidence of glint (Williams & Gaidos 2008). Near-infrared observations in the $0.8 - 1.3$ μm region at 90°–180° phase angle are best for detecting ocean glints on Earth-like planets (Robinson *et al.* 2010, Cowan *et al.* 2012.) Simultaneous attempts should also be made, if possible, to search for spectral variability in $CH_4$, $O_2$, $CO_2$, and $H_2O$ atmospheric lines denoting abundance changes due to seasonal variability as a function of planetary orbital phase. Using a space-based, stable, highly sensitive infrared telescope like JWST, seasonal variability of planetary thermal emission as a function of planetary orbital phase could also be searched for.

In sum, this procedure will be difficult and demanding, but highly rigorous and the resulting few planets to fully pass triage will be important candidates for life as we know it on an Earth 2.0-like world. We note again that Steps 5-8 can and should be re-ordered if *any* of the steps become easier, faster, or cheaper to perform because of future new techniques, facilities, or capabilities. The goal is to use the optimally cheapest and quickest way to find the "definite no's" in each step and cull the herd and reveal the true few Earth 2.0's we can robustly identify (Fig. 1).

## IV.   Examples

No logical "how-to" flowchart procedure is usefully complete without real-world concrete examples. Thus, we discuss here how our procedure would handle the well-studied planets of the Solar System, the 5 planets of the "solar system-like" 55 Cnc system, and the 7 terrestrial planets of the TRAPPIST-1 system.

Firstly, we examine our **Solar System's** planets. Assuming that accurate planetary masses and radii could be obtained for planets Venus through Neptune, Steps 1 & 2 of our logical prescription would remove all but Venus, Earth, and Mars from further consideration (using optimistic ranges for allowed planetary size and HZ; conservative ranges would exclude Venus as too hot and Mars as too small to retain an atmosphere and too cold to sustain liquid surface water at this very first step). Only the Earth, with relative atmospheric abundances of ~20% $O_2$ and ppm levels of $CH_4$, would make it to Step 8 and the search for lightcurve color variability due to land and water rotating through the observer's telescope beam. Mars would survive any triaging/culling until Step 4, where the abundance of radioactive elements and interior modeling would reveal it to be a small planet with a cold surface and a frozen lithosphere. It would also fail at Steps 5 and 6, as its atmosphere is much too tenuous to produce bluish Earth-like colors or any significant variation in transit depth with wavelength. And it would fail again at Step 7, when only 0.17% $O_2$ (Franz *et al.* 2017, Hartogh *et al.* 2010), might be detected, and methane, if detected at all, would be found at ppb to sub-ppb levels (Webster *et al.* 2015, 2018). Venus would survive until Steps 5 & 6, when its yellowish reflectance and lack of transit depth variability in multi-color lightcurves would reveal it supports an incredibly thick $H_2SO_4$ haze. Without our triage scheme, this would have been determined after only spending enough observing time to obtain Venus spectroscopy (i.e., much observing time will have been saved to use on other, more promising worlds). If one were to ignore these issues and press on to perform high-resolution spectroscopy as per Step 7, the > 1 bar of $CO_2$ (Barker & Perry 1975, Cochran *et al.* 1977) and the ppm levels of abiotically produced $O_2$ (< $8 \times 10^{-5}$, Spinrad & Richardson 1965; < $3 \times 10^{-6}$, Mills 1999) detected would eliminate it from further contention.

Secondly, we evaluate the planets in the **55 Cnc** system. We use this system as a real-life example because it is of the very few known multi-planet systems like ours, with a range of different sized planets spread out over many tens of AU from the primary star. The primary star, 55 Cancri A, has spectral type K0 IV-V, indicating an old, late-type star beginning to leave the main sequence; published age estimates for 55 Cancri A are $8.1 \pm 0.6$ Gyr (Mamajek & Hillebrand 2008) and $10.2 \pm 2.5$ Gyr (von Braun *et al.* 2011), both long enough for life as we know it to have evolved (and perhaps disappeared). The star is notably metal-rich, with a median measurement of [Fe/H] = 0.40 (Hinkel *et al.* 2014, Hinkel & Unterborn 2018); it is therefore classified as a rare "super-metal-rich" (SMR) star. It has been detected in the X-ray at a level of $\log(L_x/L_{bol})$ = -6.46 (Poppenhaeger *et al.* 2010) with typical K-star x-ray variability for a star rotating every 37.4 days (Mittag *et al.* 2017), suggesting it had enough early XUV flux to strip its planets of their primordial $H_2$/He atmospheres. However, of its 5 planets, only one —55 Cnc e— is likely rocky. With a mass ~8.3 $M_E$, this super-earth has R ~ 2 $R_{Earth}$, and thus fails the $H_2$/He atmosphere hurdle of Step 1. As well, it is located in a 0.74-day orbit at $a$ = 0.015 AU, far inside the system's HZ, also failing Step 1's tests. (There is a planet in the system's HZ, 55 Cnc f, but it is a gas giant with M > 0.16 $M_{jup}$, so it also fails at Step 1.) Thus the 55 Cnc system can be de-prioritized for further observations designed to search for biosignatures.

Thirdly, we consider the ***TRAPPIST-1*** system, where planets d, e, and f lie in the HZ of this M8V star (Gillon *et al.* 2016) that has mass 0.08 $M_\odot$, and age 7.6 Gyr (Burgasser and Mamajek 2017). TRAPPIST-1*e* and f are roughly Earth-sized, but TRAPPIST-1d is small (R = 0.78 RE) and of low enough density (~3 g $cm^{-3}$; Grimm *et al.* 2018) that only planets -1e and -1f survive past Step 1. The TRAPPIST-1 host star, spectral type M7.5 – 8.0V (Gizis *et al.* 2000; Reiners & Basri 2009, 2010), is known to be x-ray active and to flare, so e and f pass Step 2 for primary atmosphere removal. The survival of secondary atmospheres is a question in systems like this with high XUV flux (Zahnle and Catling 2017), but further characterization demands Step 2 be passed. The host star's abundances are nominally solar (but are poorly determined overall due to the difficulties inherent in detecting and measuring very late M-star atomic absorption lines) so e and f cannot be triaged at Step 3. At bulk density = 0.88 $\rho_E$ (4.84 g $cm^{-3}$) and 1.02 $\rho_E$ (5.61 g $cm^{-3}$), respectively, only TRAPPIST-1 planets c and e have densities high enough to be consistent with a rocky planet not dominated by an ocean or a H-He atmosphere, so only planet -1e survives past Step 4. (The smallish planet d would fail this step as well.)

The next steps are difficult, and so illustrate the strength of our flowchart triage technique, and the limitations of current practice. Step 5, obtaining 3-color photometry, is not feasible for the extremely low UVIS fluxes emitted from the M8V primary. This system is unlikely to be observable in reflected light in the foreseeable future due to the extremely low UVIS fluxes emitted from the M8V primary, so Step 5, obtaining 3 color photometry, is not feasible. With respect to Step 6, the system is very compact and faint, so that the first attempts to produce planetary spectra using HST/WFC3 only produced combined spectra (de Wit *et al.* 2016). A few years later, once the phasing of the individual transits was known to high accuracy, further carefully timed WFC3 observations allowed the teasing apart of 10-band WFC3 1-2 um transit spectrophotometry with $\Delta(\lambda)/\lambda$ ~4% and SNR ~2, and the determination that the -1d, -1e, and -1f data are inconsistent with cloud-free, $H_2$-dominated atmospheric models (de Wit *et al.* 2018), allowing -1e to potentially survive past Step 6. Pursuing Step 7 in the triage scheme to study -1e in detail spectroscopically will require many more (tens of nights) on the largest planned future 30m+ class ground based telescopes or 10-20 transit observations over 10+ years using the upcoming JWST (e.g., Morley *et al.* 2017, Lustig-Yaeger *et al.* 2019). For Step 8, direct imaging of any of its planets with a diffraction-limited coronagraph inner working angle of 1.22 $\lambda/D$, where $D$ is the primary aperture diameter, would require $D > 80$ m to observe in the diagnostic O2 A-band at 0.76 um, or a D > 105 m telescope to make successful 1.0 um measurements. This is likely a more optimistic estimate for the inner working angle than a real coronagraph would achieve, so even larger telescopes will be demanded. In other words, Steps 8 for -1e await bigger, more capable telescopes that are yet to be envisaged or designed, and it will likely take decades (if ever) to rigorously determine if this promising system contains an exoplanet capable of harboring Earth-like life using photometric lightcurves.

## V.    Conclusions

The procedure outlined here allows observers to start with measurements of planetary mass and radius, as well as stellar fundamental parameters (including XUV flux and elemental abundances). Then, only the most promising planets will be prioritized for the more difficult, time-consuming observations involving spectroscopy and reflectance light curves. Presumably all the planets for which our procedure is taken through to its later steps would be potentially habitable by life as we know it; but only on those with detected surface water and land could atmospheric oxygen definitely be a reliable signature of ongoing biological processes.

It is important to point out that the flowchart schema presented in this work was designed to outline straightforward logical steps for finding obvious astrobiological signatures, using examples of life as we know it that dominate the surface regions of modern Earth. It would thus implicitly miss instances of, e.g., extremophiles living in buried or isolated minority habitats, like *Halicephalobus mephisto* in South African gold mines, or archaea and bacteria in Lake Vostok (Bulat 2016). While extremophiles are important and necessary for the holistic definition of habitability, they are end-member cases that are likely to return debatable results if their remote signatures could be observed at all (Bulat 2016), and so their case must be handled carefully and at a later time, when and if observing resources for parsecs-distant Earth-sized exoplanets are capable of detecting them.

Our exercise also highlights potentially mutually exclusive selection criteria. For example, HZ exoplanets around M dwarfs are favored for atmospheric measurements, for the likelihood they transit and for their large transit depths. But optical reflectance measurements will be more easily obtained for HZ exoplanets around FGK-type stars, as M-star HZs are within the inner working angle of most telescope designs. Elemental abundances of the later M dwarfs are also difficult to obtain due to the increasing predominance of molecular vs atomic absorption features as the photosphere becomes cooler. These and other factors may need to be weighed against each other in future exoplanet characterization mission development and design.

## VI.  Acknowledgements

network sponsored by NASA's Science Mission Directorate. We acknowledge support from NExSS grant NNX15AD53G (PI Steve Desch). The authors gratefully acknowledge the support and input from colleagues across NExSS in the making of this paper, and the organizers of the Habitable Worlds Conference (Laramie, WY, November 2017), especially H. Jang-Condell, D. Gelino, and R. Kopparapu, for providing a useful forum for these interactions. We acknowledge that the research presented here uses the *Hypatia* Catalog Database, an online compilation of stellar abundance data supported by the NASA NExSS research coordination and the Vanderbilt Initiative in Data-intensive Astrophysics (VIDA) program. Part of this research was carried out at the Jet Propulsion Laboratory, California Institute of Technology, under a contract with the National Aeronautics and Space Administration. The authors thank W. Cochran and K. Stevenson for their many useful suggestions for improving this paper, and also thank an anonymous referee, whose comments greatly improved the manuscript.

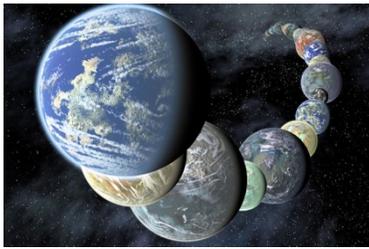

Rocky Exoplanet Oxygen Earth-Like Biosignatures

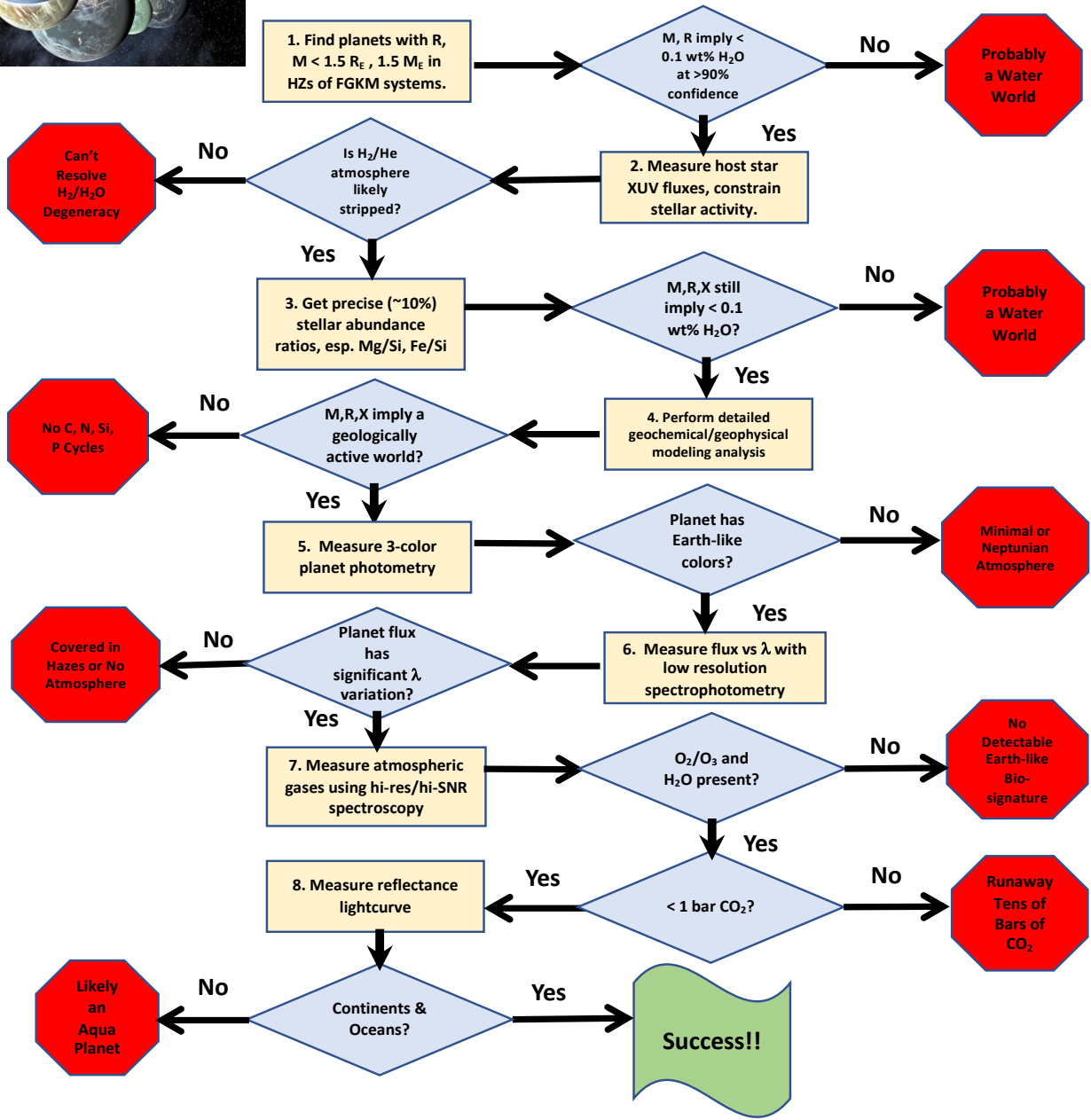

Many, many input exoplanets….

1. Find planets with R, M < 1.5 $R_E$, 1.5 $M_E$ in HZs of FGKM systems.

M, R imply < 0.1 wt% $H_2O$ at >90% confidence — **No** → Probably a Water World

**Yes**

2. Measure host star XUV fluxes, constrain stellar activity.

Is $H_2/He$ atmosphere likely stripped? — **No** → Can't Resolve $H_2/H_2O$ Degeneracy

**Yes**

3. Get precise (~10%) stellar abundance ratios, esp. Mg/Si, Fe/Si

M,R,X still imply < 0.1 wt% $H_2O$? — **No** → Probably a Water World

4. Perform detailed geochemical/geophysical modeling analysis

M,R,X imply a geologically active world? — **No** → No C, N, Si, P Cycles

**Yes**

5. Measure 3-color planet photometry

Planet has Earth-like colors? — **No** → Minimal or Neptunian Atmosphere

**Yes**

6. Measure flux vs λ with low resolution spectrophotometry

Planet flux has significant λ variation? — **No** → Covered in Hazes or No Atmosphere

**Yes**

7. Measure atmospheric gases using hi-res/hi-SNR spectroscopy

$O_2/O_3$ and $H_2O$ present? — **No** → No Detectable Earth-like Bio-signature

**Yes**

< 1 bar $CO_2$? — **No** → Runaway Tens of Bars of $CO_2$

**Yes**

8. Measure reflectance lightcurve

Continents & Oceans? — **Yes** → Success!!

**No** → Likely an Aqua Planet

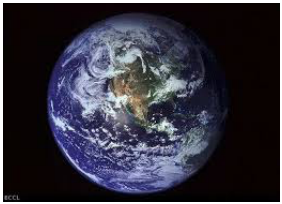

…a few surviving viable exobiology candidates

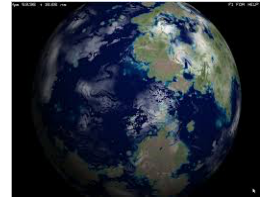

**Figure 1**: Flowchart describing an observational campaign designed to efficiently select planets for expensive observational searches for oxygen in their atmospheres so that oxygen would be a reliable biosignature (i.e., attributable to life) on those planets, if detected. The observations range from those currently being undertaken, to those requiring future ground- and space-based observations. The least time- and resource-intensive observations possible for large numbers of planets are listed first, at the top, and the most expensive and difficult measurements, possible for only a handful of exoplanets, are at the bottom, in the last part of the flowchart.